\documentclass[aps,preprint]{revtex4}

\usepackage{amsfonts}
\usepackage{amsmath}
\usepackage{amssymb}
\usepackage{graphicx}

\newcommand{\be}{\begin{equation}}
\newcommand{\ee}{\end{equation}}
\newcommand{\ba}{\begin{eqnarray}}
\newcommand{\ea}{\end{eqnarray}}

\begin{document}

\title{ Duality for symmetric second rank tensors. II. The linearized
gravitational field}
\author{H. Casini }
\affiliation{High Energy Section, \\
The Abdus Salam International Centre for Theoretical Physics, \\
I-34100 Trieste, Italy}
\author{R. Montemayor}
\affiliation{Instituto Balseiro and CAB, \\
Universidad Nacional de Cuyo and CNEA, \\
8400 Bariloche, Argentina}
\author{Luis F. Urrutia}
\affiliation{Departamento de F\'{\i}sica de Altas Energ\'{\i}as, \\
Instituto de Ciencias Nucleares, Universidad Nacional Aut\'{o}noma de M\'{e}%
xico, A.P. 70-543, M\'{e}xico D.F. 04510, M\'{e}xico}

\begin{abstract}
The construction of dual theories for linearized gravity in four dimensions
is considered. Our approach is based on the parent Lagrangian method
previously developed for the massive spin-two case, but now considered for
the zero mass case. This leads to a dual theory described in terms of a rank
two symmetric tensor, analogous to the usual gravitational field, and an
auxiliary antisymmetric field. This theory has an enlarged gauge symmetry,
but with an adequate partial gauge fixing it can be reduced to a gauge
symmetry similar to the standard one of linearized gravitation. We present
examples illustrating the general procedure and the physical interpretation
of the dual fields. The zero mass case of the massive theory dual to the
massive spin-two theory is also examined, but we show that it only contains
a spin-zero excitation.
\end{abstract}

\maketitle

\section{Introduction}

In preceding articles we have discussed the construction of dual theories
for massive fields in a Lagrangian framework, and in particular we fully
developed the case of a massive spin two theory\cite{CMUO,CMU}. The purpose
of this paper is to extend the analysis to the massless case, focusing on
the linearized gravitational field in four dimensions.

Let us recall the procedure. Starting from a second order Lagrangian, the
first step is to construct a first order Lagrangian, with a particular
structure defined by the kinetic term. It contains the derivative of the
original field times a new auxiliary variable, which corresponds to the
field strength of the original theory. The key recipe to construct the dual
theory is to introduce a point transformation in the configuration space for
the auxiliary variable which involves the completely antisymmetric tensor $%
\epsilon ^{\mu \nu \sigma \tau }$ and leads to the first order parent
Lagrangian. From the latter, both the original and the dual theories are
obtained. In fact, the equations of motion for the auxiliary variables take
us back to our starting action. Alternatively, we can eliminate the original
field from the parent Lagrangian, using its equations of motion, thus
obtaining the dual theory which is equivalent to the original one through
the transformation defined by such equations of motion.

Using the well known example of massive scalar-tensor duality we will revise
the main steps of the procedure summarized above, to point out the new
features that will appear in the massless case. Such a duality corresponds
to the equivalence between a free scalar field $\varphi $, with field
strength $f_{\mu }=\partial _{\mu }\varphi $, and an antisymmetric potential
$B_{\mu \nu }$, the Kalb-Ramond field, with field strength $H_{\mu \nu
\sigma }=\partial _{\mu }B_{\nu \sigma }+\partial _{\nu }B_{\sigma \mu
}+\partial _{\sigma }B_{\mu \nu }$\cite{FLUND,KR,DSH}. Starting from the
standard second order Lagrangian for $\varphi $ we derive a first order
Lagrangian%
\begin{equation}
L(\varphi ,L^{\mu })=L^{\mu }\partial _{\mu }\varphi -\frac{1}{2}L^{\mu
}L_{\mu }-\frac{1}{2}m^{2}\varphi ^{2}+J\varphi \ .  \label{llf}
\end{equation}%
To construct the dual theory we introduce the point transformation $L_{\mu }$
$=\epsilon _{\mu \nu \rho \sigma }H^{\nu \rho \sigma \ }$for the variable $%
L_{\mu }$, which leads to a new first order Lagrangian
\begin{equation}
L(\varphi ,H_{\nu \sigma \tau })=H_{\nu \sigma \tau }\epsilon ^{\mu \nu
\sigma \tau }\partial _{\mu }\varphi +3H_{\nu \sigma \tau }H^{\nu \sigma
\tau }-\frac{1}{2}m^{2}\varphi ^{2}+J\varphi \ .  \label{lhf}
\end{equation}%
This turns out to be the parent Lagrangian from which both theories
(original and dual) can be obtained. On the one hand, using the equation of
motion for $H_{\nu \sigma \tau }$ we get
\begin{equation}
H_{\nu \sigma \tau }(\varphi )=\frac{1}{6}\epsilon ^{\nu \sigma \tau \mu
}\partial _{\mu }\varphi \ ,  \label{HVARPHI}
\end{equation}%
which takes us back to the original second order Lagrangian for $\varphi $\
after it is substituted in Eq. (\ref{lhf}). On the other hand, we can
eliminate the field $\varphi $ from the Lagrangian (\ref{lhf}) using its own
equation of motion
\begin{equation}
m^{2}\varphi =-\partial _{\mu }\epsilon ^{\mu \nu \sigma \tau }H_{\nu \sigma
\tau }+J\ .  \label{efh}
\end{equation}%
In this way we obtain the new theory
\begin{equation}
L(H_{\nu \sigma \tau })=\frac{1}{2}\left( \epsilon ^{\mu \nu \sigma \tau
}\partial _{\mu }H_{\nu \sigma \tau }\right) ^{2}+3m^{2}H_{\nu \sigma \tau
}H^{\nu \sigma \tau }-J\epsilon ^{\mu \nu \sigma \tau }\partial _{\mu
}H_{\nu \sigma \tau }+\frac{1}{2}J^{2}\,,  \label{lhh}
\end{equation}%
which is equivalent to the original one through the transformation (\ref{efh}%
). This is a singular Lagrangian for the massive field $H_{\nu \sigma \tau }$%
, which is equivalent to a scalar field of mass $m$.

Following the same parent Lagrangian approach we have also constructed a
family of dual theories for the massive Fierz-Pauli field $h_{\mu \nu }$ in
terms of the fields $T_{(\mu \nu )\rho }$ satisfying\ $T_{(\mu \nu )\rho
}=-T_{(\nu \mu )\rho }$ and $T_{(\mu \nu )}^{\ \ \text{\ }\nu }=0$. The
cyclic identity%
\begin{equation}
T_{(\mu \nu )\rho }+T_{(\nu \rho )\mu }+T_{(\rho \mu )\nu }=0\ ,
\end{equation}%
which selects $T_{(\mu \nu )\rho }$ in the spin two irreducible
representation, was not assumed as a starting point and arose as a dynamical
result via the equations of motion\cite{CMUO,CMU}.

We now turn to the massless case. Here a very important difference appears,
which we still illustrate in the scalar field context. In this case the
equation of motion (\ref{efh}) of $\varphi $ becomes a constraint on $H_{\nu
\sigma \tau }$%
\begin{equation}
\partial _{\mu }\epsilon ^{\mu \nu \sigma \tau }H_{\nu \sigma \tau }=J\ .
\label{CONSTH}
\end{equation}%
Out of the sources, where $\partial _{\mu }\epsilon ^{\mu \nu \sigma \tau
}H_{\nu \sigma \tau }=0$, this constraint tells us that the field $H_{\nu
\sigma \tau }$ can be considered as a field strength with an associated
potential
\begin{equation}
H_{\nu \sigma \tau }=\partial _{\nu }B_{\sigma \tau }+\partial _{\sigma
}B_{\tau \nu }+\partial _{\tau }B_{\nu \sigma }\ .  \label{POTH}
\end{equation}%
In the region where $J\neq 0$ this is not valid. Hence it is not possible to
give a global solution for $B_{\sigma \tau }$ because, using Eq. (\ref{POTH}%
), the LHS of Eq. (\ref{CONSTH}) is always zero while the RHS might be non
null in a given domain. The problem is similar to that of finding the
electromagnetic potential for a magnetic monopole.

To deal with this situation we introduce a Dirac-type string singularity $%
f^{\mu }(x)$ defined by\cite{SCHWINGER}
\begin{equation}
f^{\mu }(x)=\int_{C}^{x}d\xi ^{\mu }\,\delta ^{(4)}(\xi )\ ,\qquad \partial
_{\mu }f^{\mu }(x)=\delta ^{(4)}(x)\ .
\end{equation}%
The path $C$ begins at infinity, ends up at the point $x$ and can be chosen
as a straight line if considered convenient. Thus we can write a particular
solution to Eq. (\ref{CONSTH}) as
\begin{equation}
\epsilon ^{\mu \nu \sigma \tau }\,H_{\nu \sigma \tau }(x)=\int \,(dy)f^{\mu
}(x-y)J(y)\ ,
\end{equation}%
with the general solution being
\begin{equation}
H_{\nu \sigma \tau }=\partial _{\nu }B_{\sigma \tau }+\partial _{\sigma
}B_{\tau \nu }+\partial _{\tau }B_{\nu \sigma }+\frac{1}{6}\epsilon _{\nu
\sigma \tau \rho }\int (dy)f^{\rho }(x-y)J(y)\ ,  \label{GENSOL}
\end{equation}%
in terms of the potential field and the one dimensional singular string.

The Lagrangian for the potential $B_{\alpha \beta }$ is obtained by
substituting Eq. (\ref{GENSOL}) into (\ref{lhh}), which produces the
corresponding equations of motion.

The duality transformations are expressed in terms of the following
non-local relation between the field $\varphi $, describing the original
zero-mass scalar theory, and the potential $B_{\alpha \beta }$, which is
obtained through the comparison of $H^{\mu \nu \rho }$ in Eqs. (\ref{HVARPHI}%
) and (\ref{GENSOL})
\begin{equation}
\frac{1}{6}\epsilon _{\nu \sigma \tau \mu }\partial ^{\mu }\varphi =\partial
_{\nu }B_{\sigma \tau }+\partial _{\sigma }B_{\tau \nu }+\partial _{\tau
}B_{\nu \sigma }+\frac{1}{6}\epsilon _{\nu \sigma \tau \rho }\int
(dy)f^{\rho }(x-y)J(y)\ .
\end{equation}%
Thus, in the case of massless theories the first order equation of motion
for the original variable becomes a constraint, i.e. it looks like a Bianchi
identity, which states that the dual field can now be considered as a field
strength with an associated potential plus a non-local contribution. Solving
the constraint we obtain the dual theory, in which this potential becomes
the basic field. Both theories arise from the same parent Lagrangian and
represent the same physics. This procedure strongly resembles the
electric-magnetic duality of the Maxwell theory. In fact, it is a
generalization of the well known p-form duality to arbitrary tensorial
massless fields\cite{DUAL}.

Naively one could think that another possibility to generate a massless dual
theory for the linearized gravity is to take $m=0$\ in the massive $T_{(\mu
\nu )\sigma }$\ Lagrangian of Ref. \cite{CMU}. We explore this possibility
in the Appendix, with negative results. The Dirac analysis shows that such a
theory describes only a spin zero excitation. This result corrects our
previous preliminary calculation of the number of degrees of freedom for the
massless theory reported in Refs. \cite{CMUO,CMU}, which erroneously stated
that this number was two.

The construction of dual theories is usually based on a kinematical
perspective where the basic dual fields are assigned to associated
representations of the Poincar\'{e} group\cite{LAB,LABAST}. Some dynamical
realizations of duality have also been considered in the framework of four
dimensional higher derivatives theories of gravity\cite{DESER} and in other
gravitational theories\cite{OBREGON}. Our approach is based on a Lagrangian
basis, where the auxiliary fields are not in irreducible representations to
begin with, but the ensuing Lagrangian constraints warrant that the dynamics
develops in an adequate reduced space, with a well defined spin content.

The paper is organized as follows. In the next section we apply the
dualization scheme to the Fierz-Pauli theory and obtain the dual description
in terms of two tensors, a symmetric one, $\tilde{h}_{\mu \nu }$, and an
antisymmetric one, $\omega _{\mu \nu }$. Section III contains the analysis
of the gauge symmetries of the dual theory, which clarifies the physical
meaning of the dual fields. In Section IV we consider two examples which
illustrate the construction and the effect of the dual transformations: (i)
the field describing polarized gravitational waves and (ii) the
gravitational field produced by a point mass at rest. The last section
contains a summary and some comments on the work. Finally, in the Appendix
we discuss the massless theory for the $T_{(\mu \nu )\sigma }$ field.

\section{The massless spin 2 field parent Lagrangian}

The parent Lagrangian for $m=0$ is (see Eq. (39) of Ref.\cite{CMU} with $%
a=e^{2}=1/4$)%
\begin{equation}
L=\frac{1}{8}\,T_{(\mu \nu )\sigma }T^{(\mu \nu )\sigma }+\frac{1}{4}T_{(\mu
\nu )\sigma }T^{(\mu \sigma )\nu }+\frac{1}{2}T_{(\mu \nu )\sigma }\epsilon
^{\mu \nu \alpha \beta }\partial _{\alpha }h_{\;\beta }^{\sigma }+h^{\alpha
\beta }\Theta _{\alpha \beta }\ ,  \label{plag}
\end{equation}%
where the source $\Theta _{\alpha \beta }$ is symmetric, $\Theta _{\alpha
\beta }=\Theta _{\beta \alpha }$, and conserved, $\partial ^{\alpha }\Theta
_{\alpha \beta }=0$. The field $T_{(\mu \nu )\rho }$ has zero trace, $%
T_{(\mu \nu )}{}^{\mu }=0$. From the equation of motion for $T_{(\mu \nu
)\sigma }$ we can solve $T_{(\mu \nu )\sigma }$ in terms of $h_{\;\beta
}^{\sigma }$%
\begin{equation}
\,T_{(\mu \nu )\beta }=-\epsilon _{\mu \nu }{}^{\alpha \sigma }\,\partial
_{\alpha }h_{\sigma \beta }+\epsilon _{\mu \nu \beta \lambda }\left(
\partial _{\alpha }h^{\alpha \lambda }-\partial ^{\lambda }h_{\alpha
}^{\alpha }\right) \ ,  \label{t2}
\end{equation}%
which indeed has a null trace. Plugging back the expression (\ref{t2}) in
the Lagrangian (\ref{plag}) one obtains a Lagrangian for $h^{\sigma \kappa }$%
, which is the linearized Einstein Lagrangian
\begin{equation}
\mathcal{L}=-\partial _{\mu }h^{\mu \nu }\partial _{\alpha }h_{\nu }^{\alpha
}+\frac{1}{2}\partial ^{\alpha }h_{\mu \nu }\partial _{\alpha }h^{\mu \nu
}+\partial _{\mu }h^{\mu \nu }\partial _{\nu }h_{\alpha }^{\alpha }-\frac{1}{%
2}\partial _{\alpha }h_{\mu }^{\mu }\partial ^{\alpha }h_{\nu }^{\nu
}+h^{\alpha \beta }\Theta _{\alpha \beta }\ ,  \label{hlag}
\end{equation}%
as we proved in a previous paper\cite{CMU}.

The corresponding equations of motion for $h_{\mu \nu }$ are the linearized
Einstein equations%
\begin{equation}
\partial ^{\alpha }\partial _{\alpha }h_{\mu \nu }+\partial _{\mu }\partial
_{\nu }h_{\alpha }^{\alpha }-\left( \partial _{\mu }\partial _{\alpha
}h_{\nu }^{\alpha }+\partial _{\nu }\partial _{\alpha }h_{\mu }^{\alpha
}\right) -\eta _{\mu \nu }\left( \partial ^{\alpha }\partial _{\alpha
}h_{\beta }^{\beta }-\partial _{\alpha }\partial _{\beta }h^{\alpha \beta
}\right) =\Theta _{\mu \nu }\ ,  \label{ei}
\end{equation}%
which clearly show that $h_{\;\beta }^{\sigma }$ is a spin 2 massless field.
This Lagrangian has the gauge symmetry
\begin{equation}
h_{\mu \nu }\rightarrow h_{\mu \nu }+\partial _{\mu }\varepsilon _{\nu
}+\partial _{\nu }\varepsilon _{\mu }\ .  \label{pe}
\end{equation}

On the other hand the Euler-Lagrange equation for the Lagrange multiplier $%
h^{\mu \nu }$ in (\ref{plag}) reduces to a simple constraint%
\begin{equation}
\partial _{\sigma }\left( \epsilon ^{\alpha \beta \sigma \nu }T_{(\alpha
\beta )}{}^{\mu }+\epsilon ^{\alpha \beta \sigma \mu }T_{(\alpha \beta
)}{}^{\nu }\right) =\Theta ^{\mu \nu }\ .  \label{aga}
\end{equation}%
Substituting the expression (\ref{t2}) in the above equation we recover the
equation of motion (\ref{ei}) for $h_{\mu \nu }$.

We face here a situation similar to the one already encountered for the
scalar field. From the constraint (\ref{aga}) we are able to introduce a
potential for the field $T_{(\alpha \beta )}{}^{\mu }$ only outside of the
sources. Therefore there is no global solution for a potential. In analogy
with the scalar field, we choose the particular solution ${\bar{T}}_{(\alpha
\beta )}{}^{\mu }$ of Eq. (\ref{aga}) as
\begin{equation}
\left( \epsilon ^{\alpha \beta \sigma \nu }{\bar{T}}_{(\alpha \beta
)}{}^{\mu }+\epsilon ^{\alpha \beta \sigma \mu }{\bar{T}}_{(\alpha \beta
)}{}^{\nu }\right) (x)=\int (dy)\,f^{\sigma }(x-y)\Theta ^{\mu \nu }(y)\ ,
\end{equation}%
which leads to
\begin{equation}
{\bar{T}}_{(\alpha \beta )}{}^{\mu }=-\frac{1}{6}\epsilon _{\alpha \beta
\sigma \nu }\int (dy)\,f^{\sigma }(x-y)\Theta ^{\mu \nu }(y)\ .
\end{equation}%
Next, we find the solution ${\tilde{T}}_{(\alpha \beta )}{}^{\nu }$ to the
homogeneous equation associated to (\ref{aga})
\begin{equation}
\partial _{\sigma }\left( \epsilon ^{\alpha \beta \sigma \nu }\tilde{T}%
_{(\alpha \beta )}^{\mu }+\epsilon ^{\alpha \beta \sigma \mu }\tilde{T}%
_{(\alpha \beta )}^{\nu }\right) =0\ .  \label{const}
\end{equation}%
The above equation implies that the symmetric part of the tensor $k^{\mu \nu
}=\partial _{\sigma }\epsilon ^{\alpha \beta \sigma \nu }\tilde{T}_{(\alpha
\beta )}^{\mu }$ is zero. Furthermore, this tensor has a vanishing
divergence $\partial _{\nu }k^{\mu \nu }=0$, and thus it can be written as $%
k^{\mu \nu }=\epsilon ^{\mu \nu \sigma \delta }\partial _{\sigma }A_{\delta
} $, leading to
\begin{equation}
\partial _{\sigma }\left( \epsilon ^{\alpha \beta \sigma \nu }\tilde{T}%
_{(\alpha \beta )}^{\mu }-\epsilon ^{\alpha \beta \sigma \mu }\tilde{T}%
_{(\alpha \beta )}^{\nu }\right) =2\,\epsilon ^{\mu \nu \sigma \delta
}\partial _{\sigma }A_{\delta }\ .  \label{HEQ}
\end{equation}%
In fact, Eq. (\ref{HEQ}) is a linear equation for $\tilde{T}_{(\alpha \beta
)}^{\mu }$, whose solution consists of the general solution for the
homogeneous equation plus a particular solution for the complete one. The
homogenous equation corresponding to (\ref{HEQ}) tells us that $\tilde{T}%
_{(\alpha \beta )}^{\mu }$ is a closed 2-form for each $\mu $, while a
particular solution is given by $\tilde{T}_{(\alpha \beta )}^{\mu }=\delta
_{\alpha }^{\mu }A_{\beta }-\delta _{\beta }^{\mu }A_{\alpha }$. Thus the
general solution for $\tilde{T}_{(\alpha \beta )\mu }$ in (\ref{HEQ}) is
\begin{equation}
\tilde{T}_{(\alpha \beta )\mu }=\left( \eta _{\mu \alpha }A_{\beta }-\eta
_{\mu \beta }A_{\alpha }\right) +\left( \partial _{\alpha }B_{\mu \beta
}-\partial _{\beta }B_{\mu \alpha }\right) \ ,  \label{exp}
\end{equation}%
where the tensor $B_{\mu \beta }$ is not necessarily symmetric. It can be
expressed in terms of its symmetric and antisymmetric parts, $\tilde{h}_{\mu
\nu }=\tilde{h}_{\nu \mu }$ and $\omega _{\mu \nu }=-\omega _{\nu \mu }$
respectively,%
\begin{equation}
B_{\mu \nu }=\omega _{\mu \nu }+\tilde{h}_{\mu \nu }\ .
\end{equation}%
Finally, taking into account that $\tilde{T}_{(\alpha \beta )\mu }$ must be
traceless we obtain from (\ref{exp})
\begin{equation}
A_{\alpha }=-\frac{1}{3}\left( \partial ^{\beta }\tilde{h}_{\beta \alpha
}-\partial _{\alpha }\tilde{h}_{\beta }^{\beta }+\partial ^{\beta }\omega
_{\beta \alpha }\right) \ .  \label{a}
\end{equation}%
In this way we have found the general solution for the constraint equation (%
\ref{aga}) which is
\begin{equation}
T_{(\alpha \beta )\mu }=\left( \eta _{\mu \alpha }A_{\beta }-\eta _{\mu
\beta }A_{\alpha }\right) +\left( \partial _{\alpha }B_{\mu \beta }-\partial
_{\beta }B_{\mu \alpha }\right) -\frac{1}{6}\epsilon _{\alpha \beta \sigma
\nu }\int (dy)\,f^{\sigma }(x-y)\Theta ^{\mu \nu }(y)\ .
\end{equation}%
We have obtained a description of the theory in terms of the potentials $%
\tilde{h}_{\mu \nu }$ and $\omega _{\mu \nu }$, together with a Dirac-type
string contribution.

Considering for simplicity the free field case, $\Theta _{\mu \nu }=0$, and
substituting (\ref{exp}) and (\ref{a}) into the Lagrangian (\ref{plag}) we
get
\begin{align}
L& =\frac{1}{2}\partial ^{\alpha }\tilde{h}_{\mu \nu }\partial _{\alpha }%
\tilde{h}^{\mu \nu }-\frac{2}{3}\partial _{\mu }\tilde{h}^{\mu \nu }\partial
_{\alpha }\tilde{h}_{\nu }^{\alpha }-\frac{1}{6}\partial _{\mu }\tilde{h}%
_{\alpha }^{\alpha }\partial ^{\mu }\tilde{h}_{\alpha }^{\alpha }+\frac{1}{3}%
\partial _{\mu }\tilde{h}_{\alpha }^{\alpha }\partial _{\nu }\tilde{h}^{\mu
\nu }  \notag \\
& -\frac{2}{3}\partial _{\mu }\tilde{h}^{\mu \nu }\partial ^{\alpha }\omega
_{\nu \alpha }+\frac{1}{3}\partial _{\mu }\omega ^{\nu \mu }\partial
^{\alpha }\omega _{\nu \alpha }\ .  \label{nl}
\end{align}%
In fact only the divergence of $\omega _{\nu \alpha }$ appears in the
Lagrangian, which implies that $\omega _{\nu \alpha }$ is an auxiliary
field, defined up to an arbitrary exact two form. The equations of motion
are
\begin{align}
& \partial ^{\alpha }\partial _{\alpha }\left( \tilde{h}_{\mu \nu }-\frac{2}{%
3}\eta _{\mu \nu }\tilde{h}_{\sigma }^{\sigma }\right) -\frac{2}{3}\partial
_{\alpha }\left( \partial _{\mu }\tilde{h}_{\nu }^{\alpha }+\partial _{\nu }%
\tilde{h}_{\mu }^{\alpha }\right) +\frac{1}{3}\left( \partial _{\nu
}\partial _{\mu }\tilde{h}_{\alpha }^{\alpha }+\eta _{\mu \nu }\partial
_{\alpha }\partial _{\beta }\tilde{h}^{\alpha \beta }\right)  \notag \\
& -\frac{1}{3}\partial ^{\alpha }\left( \partial _{\mu }\omega _{\nu \alpha
}+\partial _{\nu }\omega _{\mu \alpha }\right) =0\ ,  \label{x} \\
& \partial _{\beta }\left( \partial _{\mu }\tilde{h}_{\nu }^{\beta
}-\partial _{\nu }\tilde{h}_{\mu }^{\beta }\right) -\partial ^{\alpha
}\left( \partial _{\mu }\omega _{\nu \alpha }-\partial _{\nu }\omega _{\mu
\alpha }\right) =0\ .  \label{z}
\end{align}%
Eq. (\ref{z}) implies that $\partial ^{\alpha }\omega _{\nu \alpha
}-\partial _{\beta }\tilde{h}_{\nu }^{\beta }$ has zero curl. Thus
\begin{equation}
\partial ^{\alpha }\omega _{\nu \alpha }-\partial _{\beta }\tilde{h}_{\nu
}^{\beta }=\partial _{\nu }\Phi \ ,  \label{hh}
\end{equation}%
where $\Phi $ is a scalar field. The divergence of Eq. (\ref{hh}) leads to%
\begin{equation}
\partial ^{\nu }\partial _{\beta }\tilde{h}_{\nu }^{\beta }=-\partial ^{\nu
}\partial _{\nu }\Phi \ .  \label{pp}
\end{equation}%
Replacing the expression (\ref{hh}) for\textbf{\ $\partial ^{\alpha }\omega
_{\nu \alpha }$ }in (\ref{x}) we obtain
\begin{equation}
\partial ^{\alpha }\partial _{\alpha }\tilde{h}_{\mu \nu }-\partial _{\mu
}\partial _{\alpha }\tilde{h}_{\nu }^{\alpha }-\partial _{\nu }\partial
_{\alpha }\tilde{h}_{\mu }^{\alpha }+\frac{1}{3}\partial _{\nu }\partial
_{\mu }\tilde{h}_{\alpha }^{\alpha }+\frac{1}{3}\eta _{\mu \nu }\left(
\partial _{\alpha }\partial _{\beta }\tilde{h}^{\alpha \beta }-\partial
^{\beta }\partial _{\beta }\tilde{h}_{\alpha }^{\alpha }\right) -\frac{2}{3}%
\partial _{\mu }\partial _{\nu }\Phi =0\ .  \label{qqq}
\end{equation}%
Note that the trace of the the above equation does not yield $\partial ^{2}%
\tilde{h}_{\alpha }^{\alpha }-\partial ^{\alpha }\partial ^{\beta }\tilde{h}%
_{\alpha \beta }=0$, at difference with the case of the linearized
sourceless Einstein equations.

Contracting (\ref{t2}) with $\epsilon _{\kappa \lambda \sigma \alpha }{}$we
have%
\begin{equation}
\frac{1}{2}\epsilon _{\kappa \lambda \sigma \alpha }\partial ^{\kappa
}\omega ^{\sigma \lambda }\equiv D_{\alpha }=\left( \partial _{\alpha
}h_{\beta }^{\beta }-\partial _{\beta }h_{\alpha }^{\beta }\right) \ .
\label{fafa}
\end{equation}%
This equation gives the curl of the antisymmetric component, and also shows
that it is a topologically conserved current in the dual description, $%
\partial ^{\alpha }D_{\alpha }=0$. This conservation law can also be derived
from the equation of motion for $h^{\sigma \kappa }$ when the energy
momentum tensor is traceless. It expresses that in this case the scalar
curvature vanishes.

\section{Gauge symmetries}

The $T_{(\alpha \beta )}^{\mu }$ field is invariant under the following
local transformations
\begin{align}
& \delta _{\Psi }\,A_{\beta }=\partial _{\beta }\Psi \ ,\qquad \delta _{\Psi
}\,B_{\mu \beta }=\eta _{\mu \beta }\delta \Psi \ ,  \label{simpsi} \\
& \delta _{f}\,A_{\beta }=0\ ,\quad \qquad \delta _{f}\,B_{\mu \beta
}=\partial _{\mu }f_{\beta }\ ,  \label{simefe}
\end{align}%
which in terms of $\tilde{h}_{\mu \nu }$ and $\omega _{\mu \nu }$ read
\begin{align}
\delta \omega _{\mu \nu }& =-\left( \partial _{\mu }f_{\nu }-\partial _{\nu
}f_{\mu }\right) \ ,  \label{artra} \\
\delta \tilde{h}_{\mu \nu }& =\eta _{\mu \nu }\Psi +\left( \partial _{\mu
}f_{\nu }+\partial _{\nu }f_{\mu }\right) \ .  \label{deltah}
\end{align}%
The induced transformation upon the auxiliary field $\Phi $, introduced in
Eq. (\ref{hh}), is
\begin{equation}
\delta \Phi =-\Psi -2\partial _{\alpha }f^{\alpha }\ ,  \label{phi}
\end{equation}%
which shows that it is pure gauge.

The dual theory we have constructed exhibits two kinds of gauge symmetries,
one of them similar to that of the Fierz-Pauli spin two theory. Next we show
that an adequate gauge fixing for the additional $\Psi $-symmetry reduces
our theory to a standard massless spin two form. We can use the freedom in $%
\Psi $ to set
\begin{equation}
\Phi =-\tilde{h}_{\alpha }^{\alpha }\ .  \label{fijacion}
\end{equation}%
With this choice Eq. (\ref{hh}) becomes
\begin{equation}
\partial ^{\alpha }\omega _{\nu \alpha }-\partial _{\beta }\tilde{h}_{\nu
}^{\beta }=-\partial _{\nu }\tilde{h}_{\alpha }^{\alpha }\ ,  \label{hh11}
\end{equation}%
which leads to
\begin{equation}
\partial ^{\alpha }\partial ^{\beta }\tilde{h}_{\alpha \beta }-\partial ^{2}%
\tilde{h}_{\alpha }^{\alpha }=0\ .  \label{NORMA1}
\end{equation}%
As we have mentioned previously, the above equation is the trace of the
sourceless Einstein equations (\ref{ei}). Let us also remark that Eq. (\ref%
{hh11}) is invariant under the remaining gauge transformations generated by
the functions $f^{\nu }$ in Eqs. (\ref{artra}) and (\ref{deltah}). This is
because, according to Eq. (\ref{phi}), the gauge (\ref{fijacion}) fixes $%
\Psi $ without constraining $f^{\alpha }$. Thus the equations of motion (\ref%
{qqq}) can be rewritten as
\begin{equation}
\partial ^{\alpha }\partial _{\alpha }\tilde{h}_{\mu \nu }-\left( \partial
_{\mu }\partial _{\alpha }\tilde{h}_{\nu }^{\alpha }+\partial _{\nu
}\partial _{\alpha }\tilde{h}_{\mu }^{\alpha }\right) +\partial _{\nu
}\partial _{\mu }\tilde{h}_{\alpha }^{\alpha }=0\ ,
\end{equation}%
where we have explicitly used the trace condition (\ref{NORMA1}). In this
way we have recovered the linearized Einstein equations for $\tilde{h}_{\mu
\nu }$, thus describing spin two massless excitations. In particular the
gauge condition $\Phi =-\tilde{h}_{\alpha }^{\alpha }$ leads to $A_{\alpha
}=0$, according to expressions (\ref{a}) and (\ref{hh}). Therefore, we have
shown that the dual theory here obtained, described by the Lagrangian (\ref%
{nl}), is a gauge description where a $\Psi $-orbit is conformed by a set of
theories which are gauge-equivalent to the linearized Einstein theory. In
what follows we will always work in the gauge $\Phi =-{\tilde{h}}_{\alpha
}^{\alpha }$.

From Eqs. (\ref{t2}) and (\ref{exp}), the duality relation among the
potential fields is
\begin{equation}
\epsilon ^{\kappa \lambda \sigma \beta }\left( \partial _{\alpha }h_{\beta
}^{\alpha }-\partial _{\beta }h_{\alpha }^{\alpha }\right) -\epsilon
^{\kappa \lambda \alpha \beta }\,\partial _{\alpha }h_{\beta }^{\sigma
}=\left( \partial ^{\kappa }{\tilde{h}}^{\sigma \lambda }-\partial ^{\lambda
}{\tilde{h}}^{\sigma \kappa }\right) +\left( \partial ^{\kappa }\omega
^{\sigma \lambda }-\partial ^{\lambda }\omega ^{\sigma \kappa }\right) \ .
\label{DUALTR}
\end{equation}%
At this stage we can completely determine the fields of the dual theory. A
standard gauge choice in the linearized spin two theory via the functions $%
f^{\nu }$ gives ${\tilde{h}}_{\alpha \beta }$, which in turn fixes the
divergence of $\omega _{\mu \nu }$ through Eq. (\ref{hh11}), and the curl of
$\omega _{\mu \nu }$ through Eq. (\ref{fafa}), thus yielding $\omega _{\mu
\nu }$.

We can now obtain the relationship between the corresponding Riemann
tensors. Recalling its definition%
\begin{equation}
R_{\ \ \mu \nu \kappa }^{\lambda }=\frac{1}{2}\left[ \partial _{\mu }\left(
\partial _{\kappa }h_{\nu }^{\lambda }-\partial _{\nu }h_{\kappa }^{\lambda
}\right) -\partial ^{\lambda }\left( \partial _{\kappa }h_{\mu \nu
}-\partial _{\nu }h_{\mu \kappa }\right) \right] \ ,
\end{equation}%
and using Eqs. (\ref{DUALTR}) and (\ref{fafa}) we have%
\begin{equation}
\tilde{R}_{\mu \ \ }^{\ \ \sigma \lambda \kappa }=\frac{1}{2}\epsilon _{\ \
\ \ \alpha \beta }^{\kappa \lambda }\,R_{\mu }^{\ \ \sigma \alpha \beta }\ ,
\label{RIEMANN}
\end{equation}%
which exhibits the local transformation between the field strengths arising
from the non-local relation among the potentials.

Finally, it is interesting to explore the relation among the gauge
transformations in both theories. The gauge freedom due to $\varepsilon
^{\mu }$ in the original theory is mapped into gauge transformations of the
antisymmetric tensor $\omega _{\sigma \kappa }$%
\begin{equation}
\delta \omega _{\sigma \kappa }=-\,\epsilon _{\mu \nu \sigma \kappa
}{}\partial ^{\mu }\varepsilon ^{\nu }\ ,  \label{haha}
\end{equation}%
while the new gauge freedom of $\tilde{h}^{\mu \nu }$ due to $f^{\mu }$ is
independent from them. Thus the dual Lagrangian is invariant under (\ref%
{haha}) because it depends only on the divergence of $\omega _{\sigma \kappa
}$, which does not change under this gauge transformation.

\section{Examples}

In this section we discuss two examples which illustrate the construction
and effects of the proposed dual transformations. The first one refers to
the behavior of the polarization components of a gravitational wave. The
second one discusses the field produced by a point mass and shows how it is
mapped into the potentials $\omega _{0i}$ and $\tilde{h}_{0i}$, that have a
form analogous to the electromagnetic potential of a magnetic monopole.

\subsection{Gravitational waves}

In this case the gauge can be fixed using the transverse traceless gauge TT.
Working in the momentum space the gravitational field is
\begin{equation}
h_{\mu \nu }(k)=h^{+}(k)e_{\mu \nu }^{+}(k)+h^{\times }(k)\,e_{\mu \nu
}^{\times }(k)\ ,  \label{fas}
\end{equation}%
where $k^{\mu }=(k^{0},\vec{k})$, with $k_{0}=\left\vert \vec{k}\right\vert $%
, and the two possible helicities have polarization tensors
\begin{equation}
e_{\mu \nu }^{+}(k)=m_{\mu }m_{\nu }-n_{\mu }n_{\nu }\ ,\qquad e_{\mu \nu
}^{\times }(k)=m_{\mu }n_{\nu }+n_{\mu }m_{\nu }\ .
\end{equation}%
The space-like quadrivectors $m^{\mu }=(0,\hat{m})$, $n^{\mu }=(0,\hat{n})$
are such that
\begin{equation}
\hat{m}\cdot \hat{m}=\hat{n}\cdot \hat{n}=1\ ,\qquad \hat{m}\cdot \hat{n}=0\
,\qquad \hat{m}\cdot \vec{k}=\hat{n}\cdot \vec{k}=0\ .
\end{equation}%
That is to say, ${\hat{m}}$,$\,{\hat{n}}$, and ${\hat{k}}={\vec{k}}/|{\vec{k}%
}|$ form an orthonormal triad with ${\hat{n}}={\hat{m}}\times {\hat{k}}$.
Thus, the properties that define the TT gauge are $\partial ^{\mu }h_{\mu
\nu }=h_{\nu }^{\nu }=h^{0\nu }=0$. In this gauge we have $D^{\lambda }=0$
so that Eq. (\ref{fafa}) leads to
\begin{equation}
\epsilon _{\lambda \sigma \kappa \alpha }\partial ^{\lambda }\omega ^{\sigma
\kappa }=0\ .
\end{equation}

Given the field of a gravitational wave as in (\ref{fas}) we will now find
the dual fields. To begin with we also choose the TT gauge for the field ${%
\tilde{h}}_{\alpha \beta }$, which implies
\begin{equation}
\partial ^{\mu }\omega _{\mu \nu }=0\ ,
\end{equation}%
according to Eq. (\ref{hh11}). Thus we obtain $\omega _{\mu \nu }=0$. In the
chosen gauge we have the following expression for $\tilde{h}_{\mu \nu }(k)$
\begin{equation}
\tilde{h}_{\mu \nu }(k)=\tilde{h}_{+}\,e_{\mu \nu }^{+}(k)+\tilde{h}_{\times
}\,e_{\mu \nu }^{\times }(k)\ .  \label{HMUNU2}
\end{equation}%
According to Eq. (\ref{DUALTR}) the duality relations between both theories
are
\begin{equation}
{\tilde{D}}_{\kappa \sigma \lambda }=-\frac{1}{2}\epsilon _{\kappa \lambda
}{}^{\alpha \beta }\,D_{\alpha \sigma \beta }\ ,  \label{DUALTR1}
\end{equation}%
where $D^{\lambda \sigma \kappa }\equiv \partial ^{\lambda }h^{\sigma \kappa
}-\partial ^{\kappa }h^{\sigma \lambda }$. The properties ($\epsilon
^{0123}=+1$)%
\begin{equation}
\epsilon _{\alpha \beta }{}^{\mu \nu }\,k_{\mu }\,m_{\nu }=(k_{\alpha
}\,n_{\beta }-k_{\beta }\,n_{\alpha })\ ,\qquad \epsilon _{\alpha \beta
}{}^{\mu \nu }\,k_{\mu }\,n_{\nu }=-(k_{\alpha }\,m_{\beta }-k_{\beta
}\,m_{\alpha })\ ,
\end{equation}%
lead to
\begin{align}
\epsilon _{\kappa \lambda }{}^{\nu \mu }\,\left( k_{\nu }\,e_{\sigma \mu
}^{+}-k_{\mu }\,e_{\sigma \nu }^{+}\right) & =+2\left[ k_{\kappa
}\,e_{\sigma \lambda }^{\times }-k_{\lambda }\,e_{\sigma \kappa }^{\times }%
\right] \ ,  \notag \\
\epsilon _{\kappa \lambda }{}^{\nu \mu }\,\left( k_{\nu }\,e_{\sigma \mu
}^{\times }-k_{\mu }\,e_{\sigma \nu }^{\times }\right) & =-2\left[ k_{\kappa
}\,e_{\sigma \lambda }^{+}-k_{\lambda }\,e_{\sigma \kappa }^{+}\right] \ .
\label{RELBAS}
\end{align}%
The next step is to substitute Eqs. (\ref{fas}) and (\ref{HMUNU2}) in the
relation (\ref{DUALTR1}). The elements $e_{\alpha \beta }^{+}$ and $%
e_{\alpha \beta }^{\times }$ of the tensor basis in the LHS of (\ref{DUALTR1}%
) are mixed by the epsilon symbol according to Eqs. (\ref{RELBAS}), which
interchanges the labels $+$ and $\times $ of the basis tensors. Comparing
with the corresponding terms of the RHS we obtain the relations
\begin{equation}
\tilde{h}_{+}=h_{\times }\ ,\qquad \tilde{h}_{\times }=h_{+}\ .
\end{equation}%
Summarizing, the net result of the dualization procedure is to interchange
the helicity states.

\subsection{Point mass}

In the de Donder gauge $\partial ^{\nu }h_{\mu \nu }=1/2\,\partial _{\mu
}h^{\alpha }{}_{\alpha }$, the linearized gravitational field produced by a
point mass $M$ is%
\begin{equation}
h_{\mu \nu }=-\frac{2M}{r}\delta _{\mu \nu }\ ,  \label{HPM}
\end{equation}%
where the metric is $\eta _{\mu \nu }=diag(+,-,-,-)$. Note that $h_{\mu \nu
} $ is proportional to $\delta _{\mu \nu }$ and not to $\eta _{\mu \nu }$.
The trace of the gravitational field is
\begin{equation}
h=\eta ^{\mu \nu }h_{\mu \nu }=\frac{4M}{r}\ ,
\end{equation}%
so that we can rewrite $h_{\mu \nu }=-1/2\,h\,\delta _{\mu \nu }$.

The curl of $\omega ^{\sigma \lambda }$ is fixed by the original
gravitational field $h_{\mu \nu }$ through Eq. (\ref{fafa}) which yields%
\begin{equation}
-\epsilon _{\kappa \lambda \sigma \alpha }\partial ^{\kappa }\omega ^{\sigma
\lambda }=D_{\alpha }=-4\,\frac{M}{r^{3}}x_{\alpha }\ .  \label{cat}
\end{equation}%
To solve this equation we must specify the divergence of $\omega ^{\sigma
\lambda }$, which we do by choosing
\begin{equation}
\partial ^{\mu }\tilde{h}_{\mu \nu }=\partial _{\nu }\tilde{h}_{\alpha
}^{\alpha }\ ,
\end{equation}%
as the gauge in which the dual field $\tilde{h}_{\mu \nu }$ is described. In
this way Eq. (\ref{hh11}) yields
\begin{equation}
\partial ^{\mu }\,\omega _{\mu \nu }=0\ .  \label{ZDIV}
\end{equation}%
We solve Eq. (\ref{cat}) as usual. The zero divergence condition (\ref{ZDIV}%
) yields $\omega _{\mu \nu }^{H}=0$ for the regular solutions of the
corresponding homogeneous equation. As we see from Eq. (\ref{cat}) it is not
possible to give a global particular solution for $\omega ^{\sigma \lambda }$%
, because the divergence of the RHS is always zero, while the divergence of
the RHS gives $16\pi M\,\delta \left( \vec{r}\right) $. The problem is
similar to finding the electromagnetic potential for a magnetic monopole.
Using a Dirac string of the form $f^{\mu }(x)=n^{\mu }f(x)$, with the
constant vector $n^{\mu }=(0,{\hat{n}})$, we obtain that the only non zero
components of $\ \omega ^{\sigma \lambda }$ are $\omega ^{0i}=-\omega
^{i0}=H^{i}$ with
\begin{equation}
{\vec{H}}=(H^{1},H^{2},H^{3})=2M\,\frac{{\hat{n}}\times {\vec{r}}}{r(r+{\hat{%
n}}\cdot {\vec{r}})}\ ,  \label{gala}
\end{equation}%
which are singular on the negative ${\hat{n}}$ axis. The reader can verify
that we have $\nabla \cdot {\vec{H}}=0$ outside the singularity line.

Now we determine the dual field ${\tilde{h}}_{\mu \nu }$. To this end we use
the duality relations (\ref{DUALTR}) between the fields $h_{\mu \nu }$ and ${%
\tilde{h}}_{\mu \nu }$. Going back to the notation $T^{(\kappa \lambda
)\sigma }$ for the LHS of (\ref{DUALTR}) and using (\ref{HPM}) we can show
that the only non-zero contribution is
\begin{equation}
T_{(ij)0}=-\epsilon _{ij0k}{}\,\partial ^{k}h\ .
\end{equation}%
In this way, the explicit expressions for Eqs. (\ref{DUALTR}) are
\begin{eqnarray}
0 &=&\partial _{i}\tilde{h}^{00}-\partial ^{0}\tilde{h}^{0i}\ ,
\label{DUAL1} \\
0 &=&\partial _{i}\left( \tilde{h}^{j0}+\omega ^{j0}\right) -\partial ^{0}%
\tilde{h}^{ij}\ ,  \label{DUAL2} \\
0 &=&\left( \partial ^{i}\tilde{h}^{kj}-\partial ^{j}\tilde{h}^{ki}\right) \
,  \label{DUAL3} \\
\epsilon _{ijk0}\,\partial ^{k}h &=&\left( \partial ^{i}\tilde{h}%
^{0j}-\partial ^{j}\tilde{h}^{0i}\right) +\left( \partial ^{i}\omega
^{0j}-\partial ^{j}\omega ^{0i}\right) \ .  \label{DUAL4}
\end{eqnarray}%
Equation (\ref{DUAL3}), together with the symmetry of $\tilde{h}^{kj}$,
implies that $\tilde{h}^{kj}=\partial ^{k}\partial ^{j}U$ for a scalar $U$.
At this point we note that to implement the condition (\ref{ZDIV}) we have
made use of the gauge tranformations of $\omega _{\mu \nu }$, which depend
on the curl of $f_{\mu }$. We still have the gauge freedom given by $f_{\mu
}=\partial _{\mu }\Delta $, which only involves a scalar function $\Delta $.
The corresponding transformation of $\tilde{h}_{\mu \nu }$ is $\delta \tilde{%
h}_{\mu \nu }=2\partial _{\mu }\partial _{\nu }\Delta $. We can make use of
this gauge freedom to set $\tilde{h}^{jk}=0$. Eq. (\ref{DUAL2}) then gives
place to $\tilde{h}^{j0}=-\omega ^{j0}$, where we are discarding constant
solutions that do not go to zero at infinity. In consequence, the second
term in (\ref{DUAL1}) vanishes and we have $\tilde{h}^{00}=0$. Thus, from
the first three equations we get%
\begin{equation}
{\tilde{h}}^{00}=0\ ,\qquad {\tilde{h}}^{ij}=0\ ,\qquad \tilde{h}%
^{j0}=-\omega ^{j0}\ .
\end{equation}%
The remaining equation (\ref{DUAL4}) is
\begin{equation}
\partial ^{i}\omega ^{0j}-\partial ^{j}\omega ^{0i}=-\frac{1}{2}\epsilon
_{0ijk}{}\,\partial ^{k}h\ ,
\end{equation}%
which the reader can verify is only a rewriting of Eq. (\ref{cat}), which $%
\omega ^{0i}$ indeed satisfies.

Summarizing, we see that the duality introduced here maps the field of a
point source into $\omega ^{0i}$ and $\tilde{h}^{0i}$, which have the form
of the electromagnetic potential of a magnetic monopole with its
corresponding Dirac string singularity. The potential $\tilde{h}$ breaks the
rotational symmetry of the problem, but this is a gauge artifact, and as Eq.
(\ref{RIEMANN}) shows the gauge invariant quantities are symmetric under
spatial rotations.

\section{Summary and Final Comments}

Using a parent Lagrangian approach we have constructed dual theories for
linearized gravity. The starting point is the zero mass case of the parent
Lagrangian for massive spin-two theories developped in Ref. \cite{CMU}. The
equation of motion for the original field $h_{\mu \nu }$ leads to a
constraint impliying that the dual field $T_{(\mu \nu )\rho }$ can be
written as the field strength of a potential. The presence of sources
required the introduction of Dirac-type line singularities in order to have
a global solution for the potentials. The general solution for the
constraint leads to a dual description in terms of an auxiliary field $%
\omega _{\mu \nu }$, which enters only through its divergence, together with
${\tilde{h}}_{\mu \nu }$. The resulting theory has the standard gauge
symmetry of linearized gravity plus an additional $\Psi $-symmetry,
according to Eqs. (\ref{simpsi}). By an adequate gauge fixing of the latter
symmetry one recovers the Einstein equations for ${\tilde{h}}_{\mu \nu }$
together with the standard symmetries. They still affect the field $\omega
_{\mu \nu }$, (see Eqs. (\ref{artra}) and (\ref{deltah})), which becomes
determined through the gauge fixing of the gravitational fields ${h}_{\mu
\nu }$ and ${\tilde{h}}_{\mu \nu }$. In fact, such gauge fixing determines
the curl and divergence of $\omega _{\mu \nu }$ respectively, as can be seen
from Eqs. (\ref{fafa}) and (\ref{hh11}). The relation between the dual
theories is established at the level of the non-local equation (\ref{DUALTR}%
) involving the corresponding potentials ${h}_{\mu \nu }$, ${\tilde{h}}_{\mu
\nu }$ and $\omega _{\mu \nu }$. We show that this equation translates into
the somewhat expected local relation between the corresponding linearized
Riemann tensors (\ref{RIEMANN}), thus providing further evidence for the
auxiliary character of the field $\omega _{\mu \nu }$.

Two examples have been considered which illustrate the construction of the
dual theory together with the physical significance of the the dual
gravitational field ${\tilde{h}}_{\mu \nu }$. In the case of a gravitational
wave, duality just interchanges the polarizations. When considering the
field produced by a point mass, the dual configuration is a Dirac-type
string. This last example shows that the duality transformation interchanges
the role of gravitoelectric and gravitomagnetic fields, defined as
proportional to the gradient of the Newtonian potential and the curl of the $%
h^{0i}$ field respectively. Such a possibility was conjectured on the basis
of the formal similarity between Maxwell equations for the electromagnetism
and Einstein equations in the context of the PPN expansion for gravitation%
\cite{ZEE}. This duality relation also has a geometrical motivation because,
in the same way as the original Newtonian potential for a point mass is the
weak field approximation for the Schwarzschild metric, the dual field $%
\tilde{h}^{\mu \nu }$ we found is the weak field approximation for the
massless Taub-NUT metric\cite{MISNER}, which corresponds to spaces where
gravitomagnetic charges can be defined\cite{MAGNON}. Following the analogy
with the original work of Dirac on magnetic monopoles, the possibility of a
mass quantization due to the existence of a gravitomagnetic charge has also
been considered\cite{ZEE,GUO}.

Finally we have explored an alternative possibility to obtain a dual theory
for massless spin-two fields. The idea is to take the zero mass case of the
massive $T_{(\mu \nu )\rho }$ theory previously developed, which is dual to
massive Fierz-Pauli. Duality in this construction is realized in terms of
constraints that enforce a reduced phase space with the correct spin
content. It is by no means an obvious matter how these constraints and their
classification into first and second class subsets (which determines the
count of degrees of freedom) will be modified by the zero mass condition.
Hence it is difficult to know in advance which will be the spin content of
the resulting theory. We have studied this case in the Appendix, concluding
that the resulting massless theory describes spin-zero excitations. This
result corrects our previous preliminary calculation of the number of
degrees of freedom of the massless theory reported in Refs. \cite{CMUO} and
\cite{CMU}, which erroneously stated that this number was two. This
phenomena provides a clear manifestation of the van Dam-Veltman-Zakharov\cite%
{VVS} zero mass discontinuity, which leads to massless theories having
different spin content with respect to the original massive cases.We can
understand our result in terms of irreducible representations of $SO(3)$.
Since we basically start from four antisymmetric two-forms embedded in $%
T_{(\mu \nu )\rho }$, we are dealing with the product $(2,0)\times (1,0)$\
which decomposes into $(3,0)$ $+$\ $(2,1).\ $Previous results \cite%
{LABAST,CURT} indicate that the representation $(2,1)$\ carries zero degrees
of freedom, while the representation $(3,0)$\ corresponds to the Kalb-Ramond
field.

\section*{Acknowledgements}

RM\ and LFU acknowledge fruitful discussions with Antonio Garcia. This work
was partially supported by CONICET-Argentina and CONACYT-M\'{e}xico. LFU
acknowledges support from DGAPA-UNAM project IN-117000, as well as CONACYT
project 40745-F.

\appendix*

\section{The massless theory for $T_{(\protect\mu\protect\nu)\protect\sigma}$%
}

There is still another possibility we can explore when constructing massless
theories. In a preceding article we discussed dual Lagrangians for massive
spin two fields\cite{CMU}. In this approach the fields were not in
irreducible representations of the Poincar\'{e} group at the Lagrangian
level, but the Euler-Lagrange equations lead to constraints that reduced the
configuration space to the adequate representation. We can take $m=0$ in
these theories with the hope of obtaining an alternative massless spin-two
formulation. Moreover, it is not obvious \textit{a priori} how the
constraints will be modified by this choice and hence, which will be the
spin content of the resulting theory. In this Appendix we explore these
matters and show how the van Dam-Veltman-Zakharov\cite{VVS} zero mass
discontinuity, which leads to massless theories having different spin
content with respect to the original massive cases, is realized.

First we give a brief review of the well-known case of the Kalb-Ramond
field, from the perspective of the Dirac method, in order to provide a
unified description of the massive and massless cases, which clearly shows
the difference in the final counting of the true degrees of freedom.
Subsequently we present a more detailed account of the zero mass case
corresponding to the massive spin-two theory for $T_{(\alpha \beta )\mu }$
previously developed\cite{CMU}.

The massive Kalb-Ramond theory for $H_{\mu \nu \lambda }$ considered in Ref.
\cite{CURT} can be more conveniently described in terms of the field
\begin{equation}
b_{\alpha }=\frac{1}{6}\epsilon _{\alpha \mu \nu \lambda }H^{\mu \nu \lambda
}\ ,
\end{equation}%
with Lagrangian
\begin{equation}
L=\frac{1}{2}\left( \partial _{\alpha }h^{\alpha }\right) \left( \partial
_{\beta }h^{\beta }\right) -\frac{m^{2}}{2}h^{\alpha }h_{\alpha }\ .
\end{equation}%
The momenta are
\begin{equation}
\Pi _{0}=\frac{\partial L}{\partial \left( \partial ^{0}h^{0}\right) }%
=\left( \partial _{\alpha }h^{\alpha }\right) \ ,\qquad \Pi _{i}=\frac{%
\partial L}{\partial \left( \partial ^{0}h^{i}\right) }=0\ ,
\end{equation}%
leading, respectively, to
\begin{equation}
\dot{h}_{0}=\left( \Pi _{0}-\partial _{i}h^{i}\right) \ ,
\end{equation}%
together with the primary constraint $\Pi _{i}=0$. The Hamiltonian $H=\Pi
_{0}\;\dot{h}^{0}-L+\lambda ^{i}\Pi _{i}$ is
\begin{equation}
H=\frac{1}{2}\Pi _{0}^{2}+h^{i}\partial _{i}\Pi _{0}+\frac{m^{2}}{2}\left(
h^{0}h_{0}+h^{i}h_{i}\right) +\lambda ^{i}\Pi _{i}\ .
\end{equation}%
The secondary constraints are
\begin{equation}
\dot{\Pi}_{k}=0\quad \Rightarrow \quad 0=\Theta _{k}=\partial _{k}\Pi
_{0}+cm^{2}h_{k}\ .
\end{equation}%
The terciary constraints
\begin{equation}
0=\dot{\Theta}_{k}=m^{2}\left( \partial _{k}h_{0}-\lambda _{k}\right)
\end{equation}%
finalize the Dirac procedure in both cases. When $m\neq 0$ they determine
the Lagrange multiplier $\lambda _{k}$. When $m=0$ the consistency is
automatically fulfilled. Summarizing, we have the constraints
\begin{equation}
\Pi _{i}=0\ ,\qquad \Theta _{k}=\partial _{k}\Pi _{0}+cm^{2}h_{k}=0\ ,
\end{equation}%
whose classification in terms of first or second class strongly depends upon
the theory being massive or massless. In the case $m\neq 0$ the six
constraints are second class, yielding $\frac{1}{2}\left( 2\times 4-6\right)
=1$ true degrees of freedom, thus reproducing the standard scalar field.
However, the situation changes drastically in the case $m=0$. Here, the
secondary constraint $\partial _{k}\Pi _{0}=0$ reduces to only one, $\Pi
_{0}=cte$, and the remaining four constraints are first class. This leaves $%
\frac{1}{2}\left( 2\times 4-2\times 4\right) =0$ true degrees of freedom.
This is in accordance with the results of Refs. \cite{CURT} and \cite{LABAST}

Next we study the zero mass case of the dual massive spin-two theory
previously developed. Taking $m=0$ in Eq. (61) of Ref. \cite{CMU}, the
resulting Lagrangian is
\begin{equation}
\mathcal{L}=\frac{4}{9}F_{(\alpha \beta \gamma )\nu }\,F^{(\alpha \beta
\gamma )\nu }+\frac{2}{3}F_{(\alpha \beta \gamma )\nu }\,F^{(\alpha \beta
\nu )\gamma }-F_{(\alpha \beta \mu )}{}^{\mu }\,F^{(\alpha \beta \nu
)}{}_{\nu }\ ,  \label{LT0}
\end{equation}%
with
\begin{equation}
F^{(\alpha \beta \gamma )\nu }=\partial ^{\alpha }T^{\left( \beta \gamma
\right) \nu }+\partial ^{\beta }T^{\left( \gamma \alpha \right) \nu
}+\partial ^{\gamma }T^{\left( \alpha \beta \right) \nu }\ .
\end{equation}%
The Lagrangian (\ref{LT0}) dynamically fixes $T_{\ \ \ \ \ \ \beta }^{\left(
\alpha \beta \right) }=0$, and therefore it is not necessary to impose this
constraint through a Lagrange multiplier. The corresponding equations of
motion are
\begin{equation}
-4\partial _{\alpha }\,F^{\left( \alpha \nu \rho \right) \sigma }-2\partial
_{\alpha }\,F^{\left( \alpha \nu \sigma \right) \rho }+2\partial _{\alpha
}\,F^{\left( \alpha \rho \sigma \right) \nu }-2\partial _{\alpha
}\,F^{\left( \nu \rho \sigma \right) \alpha }+3\partial ^{\sigma
}F\,^{\left( \nu \rho \mu \right) }{}_{\mu }=0\ .
\end{equation}%
In order to analyze the constraints following the Dirac procedure we start
by rewriting the Lagrangian (\ref{LT0}) in terms of spatial and temporal
components
\begin{eqnarray}
\mathcal{L} &=&F_{(0ij)0}\,F^{(0ij)0}-2F^{(0ij)0}{}F_{(ijk)}^{\;\;\;%
\;k}{}-2F_{(0ij)}{}^{j}\,F^{(0ik)}{}_{k}  \notag \\
&&+\frac{4}{3}F_{(0ij)k}\,F^{(0ij)k}+\frac{4}{3}F_{(0ij)k}\,F^{(0ik)j}+\frac{%
4}{3}F_{(0ij)k}\,F^{(ijk)0}  \notag \\
&&+\frac{1}{3}F_{(ijk)l}\,F^{(ijk)l}+\frac{4}{9}F_{(ijk)0}\,F^{(ijk)0}\ .
\end{eqnarray}%
The primary constraints arising from the definition of the momenta are
\begin{eqnarray}
\Omega ^{i} &=&{\Pi }^{(ik)}{}_{k}=0\ , \\
\Gamma ^{i} &=&\Pi ^{\left( i0\right) 0}=0\ , \\
\Gamma ^{ij} &=&\Pi ^{\left( 0i\right) j}=0\ , \\
\Lambda &=&\epsilon _{ijk}\left( \Pi ^{\left( ij\right) k}-4\partial
^{i}T^{\left( jk\right) 0}\right) =0\ .
\end{eqnarray}%
The Hamiltonian is
\begin{align}
H& =\frac{1}{4}\Pi ^{(ij)0}\Pi _{(ij)0}+\frac{1}{8}\Pi _{(ij)k}\Pi
^{(ij)k}+\Pi ^{(ij)0}F_{(ijk)}^{\;\;\;\;k}  \notag \\
& -\frac{2}{3}\,F^{(ijk)0}\,F_{(ijk)0}+2T_{(j0)0}\partial _{i}\Pi
^{(ij)0}-2T_{(0j)k}\partial _{i}\Pi ^{(ij)k} \\
& +\lambda _{i}\Pi ^{(i0)0}+\lambda _{ij}\Pi ^{(0i)j}+\lambda \left( \Pi
-4\,F^{0}\right) +\mu _{i}\Pi ^{(ik)}{}_{k}\ ,
\end{align}%
with
\begin{equation}
\Pi =-\frac{1}{2}\epsilon _{ijk}\Pi ^{(ij)k}\ ,\qquad F^{0}=-\frac{1}{2}%
\epsilon _{ijk}\partial ^{i}T^{(jk)0}\ .
\end{equation}%
We see that $T_{(0j)k}$ and $T_{(0j)0}$ act as Lagrange multipliers, stating
that $\partial _{i}\Pi ^{(ij)k}=0$ and $\partial _{i}\Pi ^{(ij)0}=0$.
Therefore the degrees of freedom must be in $T_{(ij)\mu }$.

The time evolution of the primary constraints yields an additional set of
secondary constraints
\begin{align}
& \Sigma ^{i0}=\partial _{j}\Pi ^{(ji)0}\ , \\
& \Sigma ^{ij}=\partial _{k}\Pi ^{(ki)j}\ , \\
& \Sigma =\epsilon _{ijk}\left( \partial ^{i}\Pi ^{(jk)0}+\frac{4}{3}%
\partial _{r}F^{(ijk)r}\right) \ .
\end{align}%
There are no tertiary constraints.

Our set of constraints contains the first class subset%
\begin{align}
\Omega ^{i}=\Pi ^{(ik)}{}_{k}\ =0& ,\quad \rightarrow \quad 3\ , \\
\Gamma ^{i}=\Pi ^{\left( i0\right) 0}=0& ,\quad \rightarrow \quad 3\ , \\
\Gamma ^{ij}=\Pi ^{\left( 0i\right) j}=0& ,\quad \rightarrow \quad 9\ , \\
\Sigma ^{i0}=\partial _{j}\Pi ^{(ji)0}=0& ,\quad (\partial _{i}\partial
_{j}\Pi ^{(ji)0}=0)\quad \rightarrow \quad 3-1=2\ , \\
\Sigma ^{ij}=\partial _{k}\Pi ^{(ki)j}=0& ,\quad (\partial _{k}\Pi
^{(ki)}{}_{i}=0,\quad \partial _{i}\partial _{k}\Pi ^{(ki)j}=0)\quad
\rightarrow \quad 9-3-1=5\ .
\end{align}%
In parenthesis we have indicated the identities that must be substracted
when counting the number of independent constraints, which is shown to the
right of each equation. Their total number is 22. The second class subset is
\begin{align}
& \Lambda =-\frac{1}{2}\epsilon _{ijk}\left( \Pi ^{\left( ij\right) k}-\frac{%
4}{3}F^{\left( ijk\right) 0}\right) =-\frac{1}{2}\epsilon _{ijk}\left( \Pi
^{\left( ij\right) k}-4\partial ^{k}T^{\left( ij\right) 0}\right) \ , \\
& \Sigma =\epsilon _{ijk}\left( \partial ^{i}\Pi ^{(jk)0}+\frac{4}{3}%
\partial _{r}F^{(ijk)r}\right) =\epsilon _{ijk}\partial ^{i}\left( \Pi
^{(jk)0}+4\partial _{r}T^{(jk)r}\right) \ .
\end{align}%
In this way the standard count of the independent degrees of freedom $N$
gives
\begin{equation}
N=\frac{1}{2}(2\times 24-2\times 22-2)=1\ ,
\end{equation}%
showing that the massless dual theory describes a spin zero excitation.

The above count is most clearly seen in a plane wave configuration with $%
k^{\mu }=\left( k,0\ 0\ k\right) $. In this case the constraints become
\begin{align}
& \Pi ^{\left( i0\right) 0}=0\ ,\quad \Pi ^{\left( 0i\right) j}=0\ ,\quad
i=1,2,3, \\
& \Pi ^{(31)0}=\Pi ^{(32)0}=0\ , \\
& \Pi ^{(12)2}=\Pi ^{(21)1}=0\ , \\
& \Pi ^{(31)2}=\Pi ^{(32)1}=\Pi ^{(31)1}=\Pi ^{(32)2}=0\ , \\
& \Pi ^{(31)3}=\Pi ^{(32)3}=0\ ,
\end{align}%
with
\begin{equation}
T^{\left( 12\right) 0}=-\frac{1}{4k}\Pi ^{\left( 12\right) 3}\ ,\qquad \Pi
^{(12)0}=4k\,T^{(12)3}\ .
\end{equation}%
The canonical pair that remains is ($T^{(12)3}$, $\Pi ^{(12)3}$), which
means that there is only one degree of freedom, corresponding to a spin zero
field.

\end{document}